\documentclass[showpacs,preprintnumbers,prd,nofootinbib,floats,amssymb,floatfix]{revtex4}
\usepackage{amsmath}
\usepackage{amsfonts}
\usepackage{graphicx}
\usepackage{hyperref}

\usepackage{amsmath}
\usepackage{amstext}
 
\begin{document}
%\begin{flushleft}

%\documentclass{article}
%\usepackage{amsmath}
%\usepackage{amstext}
%\font\cero=cmss10 scaled 1728 \font\uno=cmssbx10 scaled 1200
%\setlength{\textwidth}{6in} \setlength{\oddsidemargin}{.3in}
%\renewcommand{\baselinestretch}{1.5}
%\setlength{\unitlength}{1ex}

\title { Hamiltonian analysis for topological  and    Yang-Mills theories  expressed as a constrained BF-like theory}
\author{ Alberto Escalante} \email{aescalan@sirio.ifuap.buap.mx}
\author{J. Angel L\'opez-Osio } \email{jalopez@sirio.ifuap.buap.mx}
 \affiliation{
 Instituto de F{\'i}sica Luis Rivera Terrazas, Benem\'erita Universidad Aut\'onoma de Puebla, (IFUAP).
   Apartado postal      J-48 72570 Puebla. Pue., M\'exico, }
 %\affiliation{ }
\begin{abstract}
The Hamiltonian analysis for  the Euler and Second-Chern  classes is performed. We show that,  in spite of the fact that  the  Second-Chern and Euler invariants give rise  to the same equations of motion, their  corresponding symplectic structures on the phase space are different,  therefore,  one can  expect different quantum formulations. In addition, the symmetries of    actions  written as a BF-like theory    that lead  to  Yang-Mills equations of motion are studied.  A close  relationship   with the results obtained in previous works  for the Second-Chern   and Euler classes is found.
\end{abstract}
\date{\today}
\preprint{}
\maketitle
%\maketitle
%\begin{enumerate}
%\item Bonne nouvelle \pause
%\item Et la mauvaise \pause
%\begin{slide}
%\item Bonne nouvelle \pause
%\item Et la mauvaise \pause

\section{ INTRODUCTION}

Nowadays $BF$ theories are  a topic of great interest in physics \cite{1,2} due to the close relation with  General Relativity, since they are background independent,  diffeomorphisms covariant  although are  devoid of local physical degrees of freedom,  \cite{3, 4, 5}.  In the literature there exist  several examples where $BF$ theories come to be relevant models, for instance in  alternative    formulations of gravity using  the MacDowell-Maunsouri approach or of Yang-Mills [YM] theory   using  Martellini's model.  MacDowell-Maunsouri formulation of gravity consists  in  breaking down   the $SO$(5) symmetry  of a  $BF$-theory from the $SO$(5)  group  to the $SO$(4),   in order to obtain   the Palatini action plus the sum of  the   Second Chern and the Euler topological invariants. Since   those topological classes  have trivial local variations  that do not contribute classically to the dynamics, one obtains essentially general relativity  \cite{5a}. Furthermore, the study of those invariants has been the subject of recent works  since  they have  a close relation with general relativity as well \cite{4}. In addition, they are expected to be related to physical observables, as for instance in the case of anomalies \cite{6,7,8,9,10,11}.
On the other hand, Martellini's model  consists in expressing  YM theory as a $BF$-like theory \cite{12}, thus, in this  so-called first-order formulation, YM
theory can be viewed as a perturbative expansion in the coupling
constant $g$ around the pure topological $BF$ theory; additionally the
$BF$ first-order formulation is {\it on shell} equivalent to the usual
(second-order) YM theory. In this context, both formulations of the
theory possess the same perturbative quantum properties \cite{13}. On the other side, there also exists  an alternative way to express  YM as a $BF$-like theory \cite{15},  and it is possible to show  again a close relation among  topological theories and physical theories. In particular, in \cite{15}   has been  analyzed  the quantum aspects  of  the  $BF$-like theory with the   abelian group $U(1)$,  and   it has been concluded  that the model   gives the same physical description than  Maxwell theory at both classical and quantum level.    \\
At the light of these facts, in this paper  we analyze the Hamiltonian formulation  for the theories discussed above. First  we perform the Hamiltonian analysis for the  Second-Chern and  the Euler invariants. We  show that,  in spite of the fact that  both  theories give rise  to the same equations of motion, their  corresponding symplectic structures are different from each other,   fact that becomes  important at both  classical and  quantum levels. It is important to mention that there exist  studies including   those invariants; in particular they play an important role at quantum level in modified versions of General Relativity  \cite{15a}. On the other hand, in \cite{16}, the canonical covariant formalism has been developed for those invariants showing that the topological classes share the Chern-Simons state within the self-dual scenario. Nevertheless,   in these  works,  the study of the symmetries   was not performed in detail, and the present work attempts to go on along these lines.
Furthermore, we perform the Hamiltonian analysis for     Martellini's model \cite{12, 13, 14} and for the action worked out  in \cite{15}. We show that both theories yield   YM equations of motion, however our Hamiltonian study shows that those theories  are not topological because there exist  physical degrees of freedom. In addition,  their  symplectic structures are different  from each other,  fact that becomes  to be relevant in the quantum treatment of both theories    just as it  happens for the Second-Chern and  Euler  classes. At the end, we show  a close relation among  these physical  actions  and  the topological  invariants discussed  at the beginning  of the paper.
\newline
%\newline
\setcounter{equation}{0} \label{c2}
 \section{ Canonical analysis for actions written as  BF-type ones:  Euler and Second-Chern Class }
 In the following two sections   we develop    Dirac's canonical analysis   for the Euler and Second-Chern topological invariants on a smaller phase space context. This means that  we shall consider as  dynamical variables  those whose  time derivatives occur in the Lagrangian. The analysis performed in this part   will be relevant for  later sections,  because we will learn that in spite of two actions give rise to   same equations of motion,  their    corresponding symplectic structures  are quite different from each other and this fact  could yield   different   quantum descriptions. Partial  results were studied in \cite{16}, however we  develop our analysis in this work by a different way extending those results.      \\
 The theories that we will study in this part  are described by  the following actions   \cite{4, 16}

 \begin{equation}
S_{SC}[A, B]=\Xi \int_M F^{IJ}\wedge B_{IJ}- \frac{1}{2} B^{IJ}\wedge B_{IJ},
\label{eqs1}
\end{equation}
and
\begin{equation}
S_{E}[A,B]=\Omega \int_M \star F^{IJ}\wedge B_{IJ}- \frac{1}{2} \star B^{IJ}\wedge B_{IJ},
\label{eqs2}
\end{equation}
where the former corresponds to the Second-Chern invariant and the later to the  Euler class. Here $ \Xi$, $\Omega$ are constants, $I, J,..=0, 1, 2, 3$  are  $SO(3,1)$ index that are raised and lowered with the Minkowski metric $\eta^{IJ}= (-1, 1, 1, 1)$,  the $\star$ product acts on internal indices namely, $\star B^{IJ}= \frac{1}{2} \epsilon^{IJ}{_{KL}}B^{KL} $,  and   we will assume that $M$ is a manifold with topology $\Sigma \times R$. \\
The equations of motion  for the action (1) are given by 
\begin{equation}
DB^{IJ}=0, \quad \quad F^{IJ}[\omega]=B^{IJ},
\label{eq3aaa}
\end{equation}
whereas for the action (2) are given by 
\begin{equation}
D\star B^{IJ}=0, \quad \quad \star F^{IJ}[\omega]= \star B^{IJ},
\label{eq4aa}
\end{equation}
after the application of the  Ò$\star$Ó operation to the equations (\ref{eq4aa}), those are reduced to (\ref{eq3aaa});    thus, we   would expect at classical level the same physical description. However, it has been remarked in \cite{ 16, 15b}  that two theories  sharing the same equations of motion,  in general do not yield  the same physical description,  in particular within the quantum context. In this way,  we need to be more careful by  carrying  out a deep  analysis of the  symmetries of the systems under study. In this respect, we will develop the Hamiltonian framework  for the actions given in (\ref{eqs1}) and (\ref{eqs2}) allowing us to    know the principal symmetries of the actions  and the relation among them. It is important to remark that  the analysis that we develop  in this section is performed by following ideas presented in \cite{15b} and  the study carry out  here   has been not reported in the literature.\\
For our  aim, it is convenient to make  the following  change of variables:  $B^i=-\frac{1}{2} \epsilon ^{i}{_{jk}} B^{jk}$, $\Upsilon^i=-\frac{1}{2}  \epsilon ^{i}{_{jk}} \omega^{jk}$, $ \pi^{a}{_{i}}= \Xi \widetilde{\eta}^{abc}B_{ibc}$, $P^{a}_{i}= \Xi \widetilde{\eta}^{abc}B_{0ibc}$,  $\tau^{i}= - \Upsilon^{i}_0$, $\Lambda^{i}=-\omega_0{^{0i}}$, $\chi_{i0a}=-2B_{ioa}$, $\varsigma_{ia}=-2 B_{0i0a}$,  $B^{i}{_{bc}}=-\frac{1}{2}  \epsilon ^{i}{_{jk}} B^{jk}{_{bc}}$,  where latin indices are raised and lowered  with the metric $\delta^{ij}=(1, 1, 1)$ and the totally anty-symmetric  Levi-Civita density of weight $+1$,  $\widetilde{\eta}^{\alpha \beta \mu \nu}$, is such that  $\widetilde{\eta}^{0123}=1$ with  $\widetilde{\eta}^{0abc}\equiv \widetilde{\eta}^{abc}$. The two-form curvature $F^{IJ}$  in terms of these variables  is given by
\begin{eqnarray}
F_{ibc}&=&  \left[  \partial_b\Upsilon_{ic}- \partial_c\Upsilon_{ib}- \epsilon_{ijk} \omega_b{^{j0}} \omega_{c0}{^{k}}+ \epsilon_{ijk} \Upsilon^j_b\Upsilon^k_c\right], \nonumber \\
F_{0ibc}&=& \left[\partial_b \omega{_{c0i}}- \partial_c \omega{_{b0i}} + \epsilon_{ijk}\omega_b{^{k}}{_{0}}\Upsilon^j_c- \epsilon_{ijk}\omega_c{^{k}}{_{0}}\Upsilon^j_b \right].
\end{eqnarray}
In this manner, using the new variables,  the Hamiltonian analysis for  the Second-Chern    and Euler invariants leads to
\begin{eqnarray}
\nonumber &S_{SC}&[\Upsilon}^i{_{a}, \pi^a_i, \omega}_a{^{0i}, P^a_i, \tau^i, \Lambda^i, \varsigma^i{_a}, \chi^i_a   ] =\int dx^0 \int_{\Sigma} dx^3 \Big \{ \dot{\Upsilon}^i{_{a}}\pi^a_i + \dot{\omega}_a{^{0i}} P^a_i  - \tau^i\left( \partial_{a}\pi^a_i+\epsilon {_{ij}}^k P^a_k \omega_a{^{0j}}+\epsilon {_{ij}}^k  \Upsilon^j_a \pi^a_k \right) \nonumber \\
&-& \Lambda^i(\partial_{a}P^a_i+\epsilon {_{ij}}^k  P^a_k \Upsilon^j_a-\epsilon {_{ij}}^k  \pi^a_k  \omega_a{^{0j}})
+ \frac{\varsigma^i{_a}}{2} \left(\pi^a_i - \Xi \widetilde{\eta}^{abc}  F_{i bc}  \right) + \frac{\chi^i_a}{2} \left( P^a_i - \Xi \widetilde{\eta}^{abc} F_{0ibc}  \right) \Big \}, \nonumber \\
\label{eq4}
\end{eqnarray}
\begin{eqnarray}
\nonumber &S _{E}&[\Upsilon}^i{_{a}, \pi^a_i, \omega}_a{^{0i}, P^a_i, \tau^i, \Lambda^i, \varsigma^i{_a}, \chi^i_a   ]= \int dx^0 \int_{\Sigma}  dx^3\Big \{ -\frac{\Omega}{\Xi} \dot{\Upsilon}^i{_{a}} P^a_i - \frac{\Omega}{\Xi} \dot{\omega}_a{^{0i}} \pi^a_i + \frac{\Omega}{\Xi} \Lambda^i ( \partial_{a}\pi^a_i+\epsilon {_{ij}}^k P^a_k \omega_a{^{0j}}   \nonumber \\ & +& \epsilon {_{ij}}^k  \Upsilon^j_a \pi^a_k )
+ \frac{\Omega}{\Xi} \tau^i(\partial_{a}P^a_i+\epsilon {_{ij}}^k  P^a_k \Upsilon^j_a- \epsilon {_{ij}}^k  \pi^a_k  \omega_a{^{0j}})
+\frac{\Omega \chi^i_a}{2\Xi}  \left(\pi^a_i -\Xi  \widetilde{\eta}^{abc}  F_{i bc}  \right) + \frac{\Omega}{2\Xi}\varsigma^i{_a} \left( P^a_i - \Xi \widetilde{\eta}^{abc} F_{0ibc}  \right) \Big \}. \nonumber \\
\label{eq5}
\end{eqnarray}
From the extended actions (\ref{eq4}) and (\ref{eq5}) we are able to identify   the corresponding symplectic  structures  for  Second Chern and Euler classes  given  by \\
Second Chern class
\begin{equation}
\{\Upsilon^i_a (x), \pi^ b_j (y) \}_{sc} = \delta^
b_a \delta ^i_ j \delta^3(x-y), \quad \quad \{\omega_a{^{0i}}(x), P^ b_j (y) \}_{sc} = \delta^
b_a \delta ^i_ j \delta^3(x-y).
\label{eq7}
\end{equation}
Euler class
\begin{equation}
\{\Upsilon^i_a (x), \frac{\Omega}{\Xi}P^ b_j (y) \}_E = \delta^
b_a \delta ^i_ j \delta^3(x-y), \quad \quad \{\omega_a{^{0i}}(x), -\frac{\Omega}{\Xi}  \pi^ b_j (y) \}_E = \delta^
b_a \delta ^i_ j \delta^3(x-y).
\label{eqs8}
\end{equation}
We  observe that  the two actions share the same dynamical variables, however, the corresponding symplectic structures are quite different from each other. In this manner, in spite of the actions (\ref{eqs1}) and  (\ref{eqs2}) giving  rise  to  the same equations of motion, in virtue of (\ref{eq7})  and (\ref{eqs8})  we expect  a different quantum description of the systems; these results confirm those reported in \cite{16} where it  was found that the Second-Chern  and the Euler classes  have different quantum states. In this way,  the action principle  presents a double role \cite{15b};  on one hand, the action  provides the equations of motion  and on the other  ones  fixes the symplectic structure, thus that the  role of the action is very  important  as matter of fact beyond  the equations of motion. Of course, there are approaches  where the equations of motion are used to determine the symplectic geometry   on  the phase  space \cite{5}, however in  that approach the phase space is not endowed with a natural or preferred symplectic structure,  this fact becomes to be  important because the freedom in the choice of symplectic structures will be relevant at the   classical and  quantum level  \cite{15b, 16}.  \\
From  (\ref{eqs1}) and (\ref{eqs2}) we are able to identify    that the actions share the following  24  constraints
\begin{eqnarray}
 \phi_i : =\partial_{a}\pi^a_i + \epsilon {_{ij}}^k P^a_k \omega_a{^{0j}}+\epsilon {_{ij}}^k  \Upsilon^j_a \pi^a_k  &\approx& 0,  \nonumber \\
 \psi_i:= \partial_{a}P^a_i + \epsilon {_{ij}}^k  P^a_k \Upsilon^j_a-\epsilon {_{ij}}^k  \pi^a_k  \omega_a{^{0j}} &\approx& 0,  \nonumber \\
\Phi^a_i:=  \left(\pi^a_i - \Xi \widetilde{\eta}^{abc} \left[ 2 \partial_b\Upsilon_{ic}- \epsilon_{ijk} \omega_b{^{j0}} \omega_{c0}{^{k}}+ \epsilon_{ijk} \Upsilon^j_b\Upsilon^k_c \right] \right)  &\approx& 0, \nonumber \\
 \Psi^a_i:= \left( P^a_i - \Xi \widetilde{\eta}^{abc} \left[2\partial_b \omega{_{c0i}} + 2\epsilon_{ijk}\omega_b{^{k}}{_{0}}\Upsilon^j_c \right] \right) &\approx& 0,
 \label{eqs9}
\end{eqnarray}
this fact  will  be relevant, because by using  the new variables defined above we  observe that the actions under study share the same constraints, however, the symplectic structures  are  different, this crucial  part was not considered in \cite{16}.  On the other side, from (\ref{eqs1}) and (\ref{eqs2}) we are able  to identify the  corresponding extended Hamiltonians   given by
\begin{eqnarray}
H_{SC_{E}} &=&\int _\Sigma \Big\{  \tau^i\left( \partial_{a}\pi^a_i+\epsilon {_{ij}}^k P^a_k \omega_a{^{0j}}+\epsilon {_{ij}}^k  \Upsilon^j_a \pi^a_k \right)
+ \Lambda^i(\partial_{a}P^a_i+\epsilon {_{ij}}^k  P^a_k \Upsilon^j_a-\epsilon {_{ij}}^k  \pi^a_k  \omega_a{^{0j}}) \nonumber \\
&-& \frac{\varsigma^i{_a}}{2} \left(  \pi^a_i -\alpha\widetilde{\eta}^{abc} \left[ 2 \partial_b\Upsilon_{ic}- \epsilon_{ijk} \omega_b{^{j0}} \omega_{c0}{^{k}}+ \epsilon_{ijk} \Upsilon^j_b\Upsilon^k_c\right]   \right)  \nonumber \\
&-&  \frac{\chi^i_a}{2} \left(  P^a_i - \alpha  \widetilde{\eta}^{abc} \left[2\partial_b \omega{_{c0i}} +2\epsilon_{ijk}\omega_b{^{k}}{_{0}}\Upsilon^j_c\right]    \right) \Big\},  \nonumber \\
H_{E_{E}}&=&\int _\Sigma \Big\{ - \frac{\Omega}{\Xi} \Lambda^i \left( \partial_{a}\pi^a_i+\epsilon {_{ij}}^k P^a_k \omega_a{^{0j}}-\epsilon {_{ij}}^k  \Upsilon^j_a \pi^a_k \right)
- \frac{\Omega}{\Xi} \tau^i(\partial_{a}P^a_i-\epsilon {_{ij}}^k  P^a_k \Upsilon^j_a- \epsilon {_{ij}}^k  \pi^a_k  \omega_a{^{0j}}) \nonumber \\
&-& \frac{\Omega \chi^i_a}{2\Xi}  \left(\pi^a_i -\Xi \widetilde{\eta}^{abc} \left[ 2 \partial_b\Upsilon_{ic}- \epsilon_{ijk} \omega_b{^{j0}} \omega_{c0}{^{k}}+ \epsilon_{ijk} \Upsilon^j_b\Upsilon^k_c\right]   \right) \nonumber \\ &-&
 \frac{\Omega}{2\Xi}\varsigma^i{_a} \left( P^a_i - \Xi  \widetilde{\eta}^{abc} \left[2\partial_b \omega{_{c0i}} + 2\epsilon_{ijk}\omega_b{^{k}}{_{0}}\Upsilon^j_c\right] \right) \Big\}.
 \label{eq10a}
\end{eqnarray}
these Hamiltonians are linear combination of constraints.  The constraints (\ref{eqs9}) are first class,  whose algebra is given by \\
Second Chern class
\begin{eqnarray}
\{\phi_{i}(x), \psi_{j}(y) \}&=& \epsilon{_{ij}}^{k}\psi_{k} \delta^3(x-y), \nonumber \\
\{\phi_{i}(x), \phi_{j}(y) \}&=& \epsilon{_{ij}}^k\phi_k \delta^3(x-y), \nonumber \\
\{\psi_{i}(x), \psi_{j}(y) \}&=& -\epsilon{_{ij}}^k\phi_k \delta^3(x-y), \nonumber \\
\{\phi_{i}(x), \Psi^a_{j}(y) \}&=& \epsilon{_{ij}}^k\Psi^a_k \delta^3(x-y), \nonumber \\
\{\psi_{i}(x), \Psi^a_{j}(y) \}&=&- \epsilon{_{ij}}^k\Phi^a_k \delta^3(x-y),  \nonumber \\
\{\Phi^a_{i}(x), \Phi^b_{j}(y) \}&=&0,  \nonumber \\
\{\phi_{i}(x), \Phi^a_{j}(y) \}&=& \epsilon{_{ij}}^k\Phi^a_k \delta^3(x-y), \nonumber \\
\{\psi_{i}(x), \Phi^a_{j}(y) \}&=&\epsilon{_{ij}}^k\Psi^a_k \delta^3(x-y),  \nonumber \\
\{\Psi^a_{i}(x), \Phi^b_{j}(y) \}&=& 0, \nonumber \\
\{\Psi^a_{i}(x), \Psi^b_{j}(y) \}&=& 0. \nonumber \\
\end{eqnarray}
Euler class
\begin{eqnarray}
\{\phi_{i}(x), \psi_{j}(y) \}&=& \frac{\Xi}{\Omega}\epsilon{_{ij}}^{k}\phi_{k} \delta^3(x-y), \nonumber \\
\{\phi_{i}(x), \phi_{j}(y) \}&=& -\frac{\Xi}{\Omega} \epsilon{_{ij}}^k\psi_k \delta^3(x-y), \nonumber \\
\{\psi_{i}(x), \psi_{j}(y) \}&=& \frac{\Xi}{\Omega}  \epsilon{_{ij}}^k\psi_k \delta^3(x-y), \nonumber \\
\{\phi_{i}(x), \Psi^a_{j}(y) \}&=& \frac{\Xi}{\Omega}  \epsilon{_{ij}}^k\Phi^a_k \delta^3(x-y), \nonumber \\
\{\psi_{i}(x), \Psi^a_{j}(y) \}&=& \frac{\Xi}{\Omega} \epsilon{_{ij}}^k\Psi^a_k \delta^3(x-y),  \nonumber \\
\{\phi_{i}(x), \Phi^a_{j}(y) \}&=& - \frac{\Xi}{\Omega}  \epsilon{_{ij}}^k\Psi^a_k \delta^3(x-y), \nonumber \\
\{\Phi^a_{i}(x), \Phi^b_{j}(y) \}&=&0, \nonumber \\
\{\psi_{i}(x), \Phi^a_{j}(y) \}&=&\frac{\Xi}{\Omega} \epsilon{_{ij}}^k\Phi^a_k \delta^3(x-y),  \nonumber \\
\{\Psi^a_{i}(x), \Phi^b_{j}(y) \}&=& 0, \nonumber \\
\{\Psi^a_{i}(x), \Psi^b_{j}(y) \}&=& 0. \nonumber \\
\end{eqnarray}
It is important to observe that    the algebra of the constraints for Second-Chern and Euler class is closed, however,   because of the symplectic structures are different,  both theories  has  an algebra with different structure,  namely;  for  Second-Chern class we see that  $\{\phi_{i}(x), \psi_{j}(y) \}$   is proportional to $\psi_{k}$,  while for Euler class    is proportional to $\phi_{k}$,  and so on with the rest of the Poisson brackets among the constraints.    On the other hand,  the extended  Hamiltonians  (\ref{eq10a}) are  a linear combination of first class constraints as  expected because of the  background independence  of the theories \cite{4,15, 16}. Furthermore,  the constraints (\ref{eqs9}) are not independent because do exist   6 reducibility conditions given by
\begin{eqnarray}
\partial_a \Phi^a_i&=&\phi_i, \nonumber \\
\partial_a \Psi^a_i&=&\psi_i. \nonumber \\
\end{eqnarray}
Thus, with all  constraints identified  we are able to carry out the counting of degrees of freedom as  follows: There are  36 canonical variables, [24-6]=18 independents first class constraints, and there are not second class constraints. Therefore  the theories  under study  are  devoid of physical degrees of freedom and correspond to  topological theories. Therefore, these results complete and extend those ones reported in  \cite{16} in the sense of    the present analysis was performed using  a different approach.\\
%\newline
%\newline
\section{ Dirac's canonical analysis for Yang-Mills theory  written as a BF-like  theory}
It has  been commented above that there exist  models where YM theory can be  written as a $BF$-like theory \cite{15}.  In the later two sections,  we   perform Dirac's canonical analysis for  two different actions, leading to  YM theory  but  with different   symplectic structures, so we will find a  similar situation  as it  was found for the   topological invariants studied above. \\
First let us  start with the following action  \cite{15}
\begin{equation}
S[A,B]= \int_M \ast  B_a \wedge B^a - 2 B_a \wedge \ast F^a ,
\label{eq14a}
\end{equation}
where $a, b, c..$ are $SU(N)$ index,   $B^a = \frac{1}{2}B^{a}_{\ \mu \nu}dx^\mu \wedge dx^\nu$ is a set of $(N^2-1)$ $SU(N)$ valued 2-forms, $ F^a= \frac{1}{2}F_{\ \mu \nu}^a dx^\mu \wedge dx^\nu$, with $F_{\ \mu \nu}^a = \partial _\mu A^a_{\nu}- \partial_\nu A^a_{\mu} + f^{a}{_{bc}}A^b_\mu A^c_\nu $ is the curvature of the connection 1-form  $A^a= A^a_{\mu}dx^\mu$.  Here, $\mu, \nu=0,1,..,3$ are spacetime indices, $x^\mu$  are the coordinates that label the points for the 4-dimensional Minkowski manifold $M$,  and $*B_{\alpha \beta}= \frac{1}{2} \varepsilon_{\alpha \beta \mu \nu}B^{\mu \nu} $ is the usual Hodge-duality operation. It is important to remark  that the   action (\ref{eq14a}) is the coupling  of two topological theories in the sense that  if we split (\ref{eq14a}), namely;  in a term $S_1[A,B]= \int_MB_a\wedge \ast F^a $  and $S_2[A,B]= \int_M \ast  B_a \wedge B^a$,  $S_1$ and  $S_2 $ are topological ones  \cite{15}. It its important to observe that    for the Euler class the star product acts on internal indices, while for the action (\ref{eq14a}) the star product  acts on space-time indices; this fact will be very important because  Euler class is a topological theory as it has been showed above, however  (\ref{eq14a})  will not be  topological anymore as it will be showed  below.   \\
So, the action on a  Minkowski background takes the following  form
\begin{equation}
S[B,A]=   \int_M \left[ \frac{1}{4}B_{\ \mu \nu}^a B_a{^{\mu \nu}} - \frac{1}{2} B^{\mu \nu }_a \left(  \partial_\mu A_{\nu}{^a}-\partial_\nu A_{\mu}{^a}+f_{\  b c}^a A_{\ \mu}^b A_{\ \nu}^c \right) \right] dx^4,
\label{eq14}
\end{equation}
the equations of motion obtained from (\ref{eq14}) read
\begin{eqnarray}
B_{\mu\nu}^{a}&=&\partial_\mu A_{\nu}{^a}-\partial_\nu A_{\mu}{^a}+f_{\  b c}^a A_{\ \mu}^b A_{\ \nu}^c, \nonumber \\
D_{\mu}B^{\mu\nu a}&=&0.
\label{eq15}
\end{eqnarray}
We can observe  that by substituting the first equation of motion in the second one,   (\ref{eq15}) is reduced to   usual YM equations of motion. We can appreciate at this level  the double role of the action (\ref{eq14}), as we have already commented above; the first one is that  the action  give us the equations of motion (\ref{eq15}), and the second one the action will fix the symplectic structure as we shall see below performing the Hamiltonian framework. \\
Thus, by performing the Hamiltonian analysis of (\ref{eq14})   we obtain
\begin{eqnarray}
&S_{E}&[A_{\mu}^{a},\Pi^{\mu }_a,B_{\mu\nu}^{a},\Pi^{\mu\nu
}_a,\lambda_{0}^{a},\lambda^{a},u_{i}^{a},u_{0i}^{a},u_{ij}^{a},v_{ij}^{a}]=\int
d^{4}x(\dot{A}_{\mu}^{a}\Pi^{\mu}_a+\dot{B}_{\mu\nu}^{a}\Pi^{\mu\nu}_a-\frac{1}{2}\Pi^{i}_a\Pi_{i}^{a} +\frac{1}{4}B_{ij}^{a}B^{ij}_a \nonumber\\
&+&A_{0}^{a}
D_{i}\Pi^{i}_a-\frac{1}{2}B^{ij}_{a}F_{ij}^{a}
-\lambda_{0}^{a}\gamma^{0}_a-\lambda^{a}\gamma_{a}
-u_{i}^{a}\chi^{i}_a-u_{0i}^{a}\chi^{0i}_a-u_{ij}^{a}\chi^{ij}_a-v_{ij}^{a}\phi^{ij}_a),
\label{eq16}
\end{eqnarray}
here $(\Pi^{\alpha}_a, \Pi^{\alpha \beta}_a) $ are canonically conjugate to $(A^a_{\alpha}, B^a_{\alpha \beta})$;  $i,j,k..=1,  2, 3$,    and  $D_i\pi^{i}_a= \partial_i \pi^{i}_a + f_{abc}A_i^b \pi^{ic}$. \\
From (\ref{eq16}) can be identified the following symplectic structure
\begin{equation}\begin{split}
\{A_{\alpha}^{a}(x),\Pi^{\mu
}_b (y)\}&=\delta^{\mu}_{\alpha}\delta^{a}_b\delta^{3}(x-y),\\
\{B_{\alpha\beta}^{a}(x),\Pi^{\mu\nu
}_b(y)\}&=\frac{1}{2}\left(\delta_{\alpha}^{\mu}\delta_{\beta}^{\nu}-\delta_{\beta}^{\mu}\delta_{\alpha}^{\nu}\right)
\delta^{a}_b\delta^{3}(x-y),
\end{split}
\label{eq18a}
\end{equation}
and  $\lambda_{0}^{a}$, $\lambda^{a}$ are Lagrange multipliers enforcing the following  $2(N^2-1)$ first class constraints
\begin{equation} \label{fc bf}\begin{split}
\gamma^{0}_a&=\Pi^{0}_a\approx 0,\\
 \gamma_{a}&=D_{i}\Pi^{i}_ a+2 f{_{abc}}B_{0i}^{b}\Pi^{0ic}+f{_{abc}}B_{ij}^{b}\Pi^{ijc}\approx 0, \\
\end{split}
\end{equation}
$u_{i}^{a}$,  $u^a_{0i}$,  $u^a_{ij}$,  $v^a_{ij}$ are Lagrange multipliers enforcing the following $12(N^2-1)$ second class constraints
\begin{equation}\label{sc bf}\begin{split}
\chi^{i}_a&=\Pi^{i}_a+B^{0i}_a\approx 0,\\
\chi^{0i}_a&=\Pi^{0i}_ a\approx 0,\\
\chi^{ij}_a&=\Pi^{ij}_a\approx 0,\\
\phi^{ij}_a&=B^{ij}_a-F^{ij}_a \approx 0.
\end{split}
\end{equation}
Therefore, the counting of degrees of freedom is carry out as follows; there are  $20(N^{2}-1)$ phase space variables,
$2(N^{2}-1)$ independent first class constraints and $12(N^{2}-1)$
second class constraints, thus the theory given in (\ref{eq16}) has
$2(N^{2}-1)$ degrees of freedom, corresponding  to  the number of degrees of freedom for YM theory. \\
From the action  (\ref{eq16}) we  also identify the extended Hamiltonian for the theory
\begin{equation}
H_E=\frac{1}{2}\Pi^{i}_a\Pi_{i}^{a} -\frac{1}{4}B_{ij}^{a}B^{ij}_a
-A_{0}^{a}
D_{i}\Pi^i_a+\frac{1}{2}B^{ij}_{a}F_{ij}^{a}
-\lambda_{0}^{a}\gamma^{0}_a-\lambda^{a}\gamma_{a}.
\label{eq19}
\end{equation}
Of course, the   difference between  the Hamiltonian (\ref{eq19}) and  the Hamiltonians (\ref{eq10a}),   is that (\ref{eq19}) is not linear combination of constraints; this fact is due to the action (\ref{eq14a}) is not background independent. Furthermore, if the second class constraints (\ref{sc bf}) are considered as strong equations, we recover the standard YM Hamiltonian.  \\
With the constraints identified,  we are able to calculate the gauge transformations  on the phase space. For this aim we define the gauge generator in the following  form
\begin{equation}
G=\int_{\Sigma}\left[D_{0}\epsilon_{0}^{a}\gamma^0_{a}+\epsilon^{a}\gamma_{a}\right]d^{3}x,
\end{equation}
so, the gauge transformations are given by
\begin{eqnarray}
\{ A_{\alpha}^a, G \}= -D_\alpha\epsilon^a, \nonumber \\
\{ \Pi^{\alpha}_a, G \}= - f_{ba}{^{c}}\Pi^\alpha_c \epsilon^b,
 \label{eq89}
\end{eqnarray}
or
\begin{eqnarray}
A^a_\alpha&\rightarrow& A^a_\alpha-  D_\alpha \epsilon^a , \nonumber \\
\Pi_a ^\alpha &\rightarrow& \Pi_a ^\alpha - \epsilon^b f_{ba}{^{c}} \Pi_c^\alpha .
\label{eq22b}
\end{eqnarray}
We finish this section with some remarks. On one hand, it is important to perform the quantum treatment of the action (\ref{eq14a})  and to find  the differences respect to the standard YM theory; this subject of study is already in progress \cite{15}.   On the other, we can also introduce a constant $g^2$ in the second term of the action (\ref{eq14a}) namely,  $S[A,B]= \int_M g^2 \ast  B_a \wedge B^a -2B_a\wedge \ast F^a $,  thus,  we are able to  analyze the perturbative behavior in the constant $g$ around the second   term   $\int_M B_a\wedge \ast F^a $ which corresponds to be topological one, and then comparing  this behavior with the results obtained in \cite{14} for Martellini's model. On the other side, in the following section we will analyze the relation between  the action (\ref{eq14a}) and Martellini's model because both actions yield  {\it on shell} to YM theory, however we shall see that their  corresponding symplectic structures are different jus as in the case for  topological invariants studied above.
\section{Dirac's canonical analysis for Martellini's model}
It is well-know  in the literature that there exists a different  model  to (\ref{eq14a}),    with the particularity that  its  equations of motion also yield  YM equations; that one is called Martellini's model \cite{12,13,14}. As it has been comment above, Martellini's model  is a deformation of a $BF$ topological field theory where it is possible show that  gives
the first order formulation of YM theory,   and it has been showed that the
standard $uv$-behaviour is recovered \cite{14}, and  new non local observables can be defined, which are
related to the phase structure of the theory.\\
For these reasons, in this section we shall perform the Hamiltonian analysis for Martellini`s model which is absent  the literature,  and we will compare the results obtained in this part with those found in Sec. III, all this part will be  developed with in  the aim to observe if there exists  a similar situation just as for Second-Chern and Euler invariants. \\
Our starting  point is   the following action \cite{12, 13, 14}
 \begin{equation}
 S[A,B]=   \int_M \mathbf{i} B_a \wedge F^a+ \frac{g^2}{4} B_a\wedge \ast B^a,
 \label{eq20a}
 \end{equation}
where  $\mathbf{i}$ is the complex number,  $B^a$,  $F^a$ and $*B$ are defined as in Sec. III, $g$ is a   coupling constant. We can observe that the action (\ref{eq14a})   and (\ref{eq20a}) share the same dynamical variables, however they are different, in  (\ref{eq14a})  the star product acts in both terms of the action, while in    (\ref{eq20a}) it does not;  something similar  occurs  for the topological invariants. Furthermore, (\ref{eq20a}) has an imaginary factor $\mathbf{i}$ which is relevant in the path integral formulation.    \\
Hence,  on a Minkowski space-time the action takes the form
\begin{equation}
S[A,B]= \int_M \left[ \frac{\mathbf{i}}{4}\epsilon^{\mu \nu \alpha \beta} F_{a \mu \nu}B_{\ \alpha \beta}^a+\frac{g^2}{4} B_{a \mu \nu}B^{a \mu \nu}  \right] dx^4.
\label{eq24a}
\end{equation}
The equations of motion obtained from (\ref{eq24a}) read
\begin{eqnarray}
\frac{\mathbf{i}}{4}\epsilon^{\mu \nu \alpha \beta} F^a_{ \mu \nu}&=& - \frac{g^2}{2} B^{a \alpha \beta}, \nonumber \\
D_{\beta}(\epsilon^{\mu \nu \alpha \beta} B_{a \mu \nu}) &=&0,
\label{eq26c}
\end{eqnarray}
where the derivation $D_\beta$ is defined as above. We  observe that  the action  (\ref{eq20a}) yields on shell   YM  theory  (just as the  action (\ref{eq14a})). \\
 Thus,  the Hamiltonian analysis of the action (\ref{eq24a}) leads to
\begin{align}
S_{E}& [A_{\mu}^{a},\Pi^{\mu }_a,B_{\mu\nu}^{a},\Pi^{\mu\nu
}_a,\lambda_{0}^{a},\lambda^{a},u_{i}^{a},u_{0i}^{a},u_{ij}^{a},v_{0i}^{a}]=\int
d^{4}x(\dot{A}_{\mu}^{a}\Pi^{\mu }_a+\dot{B}_{\mu\nu}^{a}\Pi^{\mu\nu
}_a-\frac{1}{2}\Pi^{i}_a\Pi_{i}^a + 2g^{2}B_{0i}^{a}B^{0i}_a \nonumber\\
+A_{0}^{a}& D_{i}\Pi^{i}_a+\mathbf{i}\eta^{ijk}B_{a0i}F_{jk}^{a}
-\lambda_{0}^{a}\gamma^{0}_a-\lambda^{a}\gamma_{a}
-u_{i}^{a}
\phi^{i}_a-u_{0i}^{a}\phi^{0i}_a-u_{ij}^{a}\phi^{ij}_a-v_{0i}^{a}\psi^{0i}_a),
\label{eq23}
\end{align}
where $(\Pi^{\alpha}_a, \Pi^{\alpha \beta}_a) $ are canonically conjugate to $(A^a_{\alpha},  B^a_{\alpha \beta})$, and  we are able to identify the symplectic structure
\begin{equation}\begin{split}
\{A_{\alpha}^{a}(x),\Pi^{\mu
}_b (y)\}&=\delta^{\mu}_{\alpha}\delta^{a}_b\delta^{3}(x-y),\\
\{B_{\alpha\beta}^{a}(x),\Pi^{\mu\nu
}_b(y)\}&=\frac{1}{2}\left(\delta_{\alpha}^{\mu}\delta_{\beta}^{\nu}-\delta_{\beta}^{\mu}\delta_{\alpha}^{\nu}\right)
\delta^{a}_b\delta^{3}(x-y).
\end{split}
\label{eq23a}
\end{equation}
On the other hand,  $\lambda_{0}^{a}$, $\lambda^{a}$ are Lagrange multipliers enforcing the following  $2(N^2-1)$ first class constraints
\begin{equation} \label{Mfcc}\begin{split}
\gamma^{0}_a&=\Pi^{0}_a\approx 0,\\
 \gamma_{a}&=D_{i}\Pi^{i}_a+2f_{abc}B_{0i}^{b}\Pi^{0ic}+f_{abc}B_{ij}^{b}\Pi^{ijc}\approx 0,
\end{split}
\end{equation}
and $u_{i}^{a}$, $u_{0i}^{a}$, $u_{ij}^{a}$, $v_{0i}^{a}$ are Lagrange multipliers enforcing the following $12(N^{2}-1)$ second class constraints

\begin{equation}\label{Mscc}\begin{split}
\phi^{i}_a&=\Pi^{i}_a-\mathbf{i}\eta^{ijk}B_{jka}\approx 0,\\
\phi^{0i}_a&=\Pi^{0i}_a\approx 0,\\
\phi^{ij}_a&=\Pi^{ij}_a\approx 0,\\
\psi^{0i}_a&=2g^{2}B^{0i}_a+\frac{\mathbf{i}}{2}\eta^{ijk}F_{a jk}\approx 0.
\end{split}
\end{equation}
It is important to remark that in (\ref{Mfcc}) the second  first class constraint  corresponds to the Gauss constraint for this theory;  on the other side,  in virtue of (\ref{Mscc}) the symplectic structure (\ref{eq18a}) and  (\ref{eq23a}) are different  since  in  (\ref{eq14}) the definition of the momenta is $\Pi^{i}_a=- B^{0i}_a$, while for (\ref{eq24a}) is $\Pi^{i}_a= \mathbf{i}\eta^{ijk}B_{jka}$,   and this fact will be  important in the quantum treatment. So, in spite of both actions  (\ref{eq14}) and (\ref{eq20a})  yield {\it on shell}   YM theory and  the fact that their  corresponding  symplectic structures are different from each other, we   could expect    a similar situation just as   Second-Chern and Euler invariants studied in Sect. III,   where the quantum theories are different. However,  the quantum study of (\ref{eq14}) is still in progress and we can not say yet if at quantum level the action (\ref{eq14}) corresponds to YM theory; nevertheless,    for an  abelian group  the action (\ref{eq14}) is equivalent to  Maxwell theory at classical and quantum level \cite{15}.  \\
On the other hand,  the  extended Hamiltonian for the theory (\ref{eq24a}) is given by
\begin{equation}
H_E=\frac{1}{2}\Pi^{i}_a\Pi_{i}^a - 2g^{2}B_{0i}^{a}B^{0i}_a
-A_{0}^{a} D_{i}\Pi^{i}_a-\mathbf{i}\eta^{ijk}B_{a0i}F_{jk}^{a}
-\lambda_{0}^{a}\gamma^{0}_a-\lambda^{a}\gamma_{a}.
\end{equation}	
With all this  information at hand, we are able to carry out the counting of physical degrees  of freedom as follows;  There are  $20(N^{2}-1)$ phase space variables,
$2(N^{2}-1)$ independent first class constraints and $12(N^{2}-1)$
second class constraints, thus the theory given in (\ref{eq20a}) has
$2(N^{2}-1)$ degrees of freedom.  \\
 Now,  it is straightforward to prove that the gauge transformations on the phase space for the theory under study   are those for YM theory. For this aim  we define the gauge generator in the  form
\begin{equation}
G=\int_{\Sigma}\left[D_{0}\epsilon_{0}^{a}\gamma^{0}_a+\epsilon^{a}\gamma_{a}\right]d^{3}x,
\end{equation}
hence, the gauge transformations are given by
\begin{eqnarray}
\{ A_{\alpha}^a, G \}= -D_\alpha\epsilon^a, \nonumber \\
\{ \Pi^{\alpha}_a, G \}= - f_{ca}{^{b}}\Pi^\alpha_b \epsilon^c,
 \label{eq89}
\end{eqnarray}
or
\begin{eqnarray}
A^a_\alpha&\rightarrow& A^a_\alpha-  D_\alpha \epsilon^a , \nonumber \\
\Pi_a ^\alpha &\rightarrow& \Pi_a ^\alpha - \epsilon^c f_{ca}{^{b}} \Pi_b^\alpha .
\label{eq22b}
\end{eqnarray}
Therefore,  we have presented the Hamiltonian study for  two actions  that give rise  to YM equations of motion,  nevertheless,   their  respective symplectic structures are different. The actions  have  a close relation with topological theories, and  we believe  that  these facts will be relevant  in the quantum scenario. The analysis  presented in this paper has been performed at classical level and the quantum approach will be  reported  in forthcoming works  \cite{15},  expecting  to confirm a similar situation as was  found for  the Second-Chern and the Euler  invariants \cite{16}.
%%%%%%%%%%%%%%%%%%%%%%%%%%%%%%%%%%%%
\newline
\newline
%\section{ Conclusions and prospects}
\noindent \textbf{Acknowledgements}\\[1ex]
This work was supported by CONACyT  M\'exico under grant 95560 and Sistema Nacional de Investigadores (SNI).  J. Angel L\'opez-Osio thanks  Vicerrector{\'i}a de investigaci\'on y estudios de posgrado (BUAP) for support.   \\

\end{document}